\documentclass[12pt,a4paper]{article}
\usepackage[english]{babel}
\usepackage[utf8]{inputenc}
\usepackage[dvips]{graphics}
\usepackage{longtable}
\usepackage{lscape}
\usepackage{multirow}
\catcode`\{=1
\catcode`\}=2
\def\lref#1{{\tt \textsuperscript{[#1]}}}

\title{A Domain-specific Language for High-reliability Software used in the JUICE SWI Instrument -- The hO Language Manual}
\author{Felix Winkelmann\footnote{e-mail: {\it winkelmann@mps.mpg.de}} ,~
	Oskar Schirmer\footnote{e-mail: {\it oskar.schirmer@mps.mpg.de}}}
\date{Max-Planck-Institut f\"{u}r Sonnensystemforschung \\
	G\"{o}ttingen, 2017-09-11}
\begin{document}

\maketitle
\vfill

\section*{Abstract}

{\it hO} is a custom restricted dialect of Oberon,
developed at the Max-Planck Institute for Solar System Research in G\"{o}ttingen
and used in the SWI flight software for the JUICE mission.
{\it hO} is applied to reduce the possibility of
syntactically valid
but incorrect code, provide better means of statically analyzing
source code, is more readable than C and gives syntactic
support for the software architecture used in the SWI instrument
software. By using a higher-level, application-specific notation
a whole range of possible errors is eliminated and source code size is
reduced,
while making the code itself easier to understand, review and
analyze.

\vfill
\newpage

\tableofcontents
\newpage

\section{Introduction}

This document describes the {\it hO} language, a dialect of Oberon\lref{26},
which is used as a replacement for
C\lref{17} in the implementation of the
SWI\lref{14}
flight software for the JUICE\lref{8}
mission.

The SWI flight software was initially planned to be implemented in C
which is a de-facto standard for most software written for current
and future ESA missions, as C is widely available and relatively well known.
C is also directly supported by the main hardware platform\lref{10} that is used
throughout the JUICE project in the form of the Gaisler BCC toolchain\lref{11}
and debugging utilities.

For embedded systems in constrained environments, C is a common choice
as an implementation language, but carries with it a considerable class
of hazards regarding memory-safety and the handling of exceptional
situations. C's history as a systems programming language puts an
emphasis on full control over hardware resources and maximum efficiency,
but intentionally removes run-time checks and provides means to circumvent
the type-system in various ways, for example by using type-casts and
unions.  In an environment where high-reliability and fault tolerance
are at a premium, as is the case in space projects, the flexibility
provided by C can be a liability, because error-checking and fault recovery
are entirely the responsibility of the programmer. Moreover, only very
low-level and general control structures are available which, when used,
need careful attention to avoid out-of-bound errors, non-terminating
iterations or unhandled cases when discriminating among a set of values.
Additionally, unsafe pointer-handling makes it hard to ensure the absence
of bugs and results in hard to read code, incurring additional time for
reviews and more extensive levels of debugging\lref{23}\lref{2}.

Therefore we decided to use a custom, domain-specific language that is
easier to parse -- both for humans and tools -- and that provides direct
syntactic support for the architecture that we use in the SWI flight
software. We briefly considered existing languages targeted for embedded
high-reliability use:

Most obviously, Ada\lref{16} may be
considered the prime choice for software development in this area,
but a lack of programming experience in this language and
the existing implementations and runtime-systems made us look
for simpler and more lightweight solutions.

Another option would be Forth\lref{21}, which is already widely
used in space applications, is very lightweight and extremely
flexible. The problem here was that this would not have permitted
us to use the Gaisler C toolchain and the greatly different
philosophy of Forth would have required a complete redesign
of the already existing prototypical C codebase that we used
for evaluating the architecture.

Finally, there are more recent, mostly academic attempts at
designing languages for high-reliability software like Ivory\lref{4},
but a lack of good documentation, a somewhat unstable implementation,
and the very obscure syntax made us reject this option.

As both authors have experience in language
design and implementation\lref{22}\lref{25}, we decided to create a notation
and translator of our own, tailored to the software architecture
used in the SWI project.

We still use C as a target language, to avoid the intricacies of
direct native code generation and take advantage of existing
optimization technology. Moreover, using the vendors toolchain
takes advantage of workarounds for hardware bugs in the board that
is used for testing the DPU software\lref{12}.

\section{Architecture}

As the architecture of the SWI flight software had considerable influence
on the language design of hO, a short description of the structure of the
application is given here.

The SWI application employs a {\it cyclic executive}\lref{1}\lref{18} to schedule
repeating and concurrent tasks. Functionality is separated into {\it modules}
which are executed one after the other, repeatedly. Modules can be
separated into the following groups:

\begin{description}

\item a) Modules responsible for sub-units of the instrument (spectrometers, motors, oscillator).

\item b) Modules processing specific hardware (SpaceWire, SPI, GPIO, ADC/DAC).

\item c) Modules processing incoming telecommand- (TC) and outgoing telemetry packets (TM).

\item d) Modules performing checks of the overall status of the system as a whole and
reporting the status to the on-board computer and ground (memory, safety checks,
house keeping, watchdog).

\item e) Modules processing higher-level funtionality, that is, sequences of operations specific
to the instrument (scripts).

\end{description}

Long-running tasks have to be split up manually to ensure a minimum latency between
module executions, but the mininum latency for a complete
execution cycle can be determined statically, and the system is fully
deterministic: tasks are interrupted at well known and safe locations.

SWI does not use an operating system. The functional and performance
requirements for the software are moderate and most hardware interfaces
are SWI-specific. Therefore the additional complexity and
size of an operating system is not justified. Moreover, having complete
control of the system and being able to understand all parts of it in detail
gives us the confidence that the number of errors in the final
product is minimized.

Modules communicate by zero-copy message passing: each module has
a public data area, accessible from other modules, and a private data
area for internal use. A special structure named {\it port} holds a single message
during the transmission from module A to module B. Once a message
is processed, the port is emptied, the memory occupying the message
is returned to the memory management subsystem and the port can
receive a fresh message. Message buffers have a fixed size (just enough
to hold the maximum telecommand/telemetry message packet size). Since
nearly all messages are TC/TM packets or are derived from those, no
copying is involved, as long as we ensure messages conform to the
TC/TM layout. Internal messages for programmatic control of some parts of the
system ({\it scripting}) use TC packets as messages, for example, and thus
do not need special handling of packets that originated from ground control
or from an internal source.

Memory from messages is dynamically allocated from a pool of fixed-size
message blocks, which makes the memory management very simple
and transparent. Unused memory is reclaimed as early as possible and
messages can not be queued, avoiding the potential of ports filling
up with unhandled messages and subsequent exhaustion of the available
memory pool.

\section{Language}

hO is based on Oberon\lref{26}, a Pascal/Modula dialect developed by Niklaus
Wirth. Oberon has the advantage of being syntactically simple and easy
to understand, it has a simple runtime model and a minimal runtime
system. hO further simplifies Oberon by removing support for strings,
real numbers, procedure types, multi-dimensional arrays, derived record types, sets,
{\tt WITH} and most loop types. Dynamic memory management has been
removed, with the exception of {\tt PORT} types.

By choosing the simpler and stricter notation of Oberon instead of a C
based syntax, many of the classic pitfalls in C programming are
avoided. Specifically, of the 142 rules specified in the MISRA-C 2004
guidelines\lref{20}, a total of 116 rules are automatically
enforced by the syntactic rules of the hO language and properties
of the generated code and runtime system.


Similarly, only one out of the ten rules proposed with
``The Power of Ten''\lref{15}
continues to play a role when switching from C to hO,
while the other nine rules are obsolete.
Moreover, while translating a fully functional prototype
of the software from C to hO,
the amount of source code did not increase,
instead decreasing slightly from 3336 to 2853 LOC.

Extensions to Oberon that have been added to reflect the SWI software
architecture are: syntax for state machines, a {\tt PORT} type and built-in
procedures for message passing, extensions for {\it Design-by-Contract}\lref{19} and
some low-level utilities for direct byte-level memory access. The {\tt MODULE}
construct has been modified to simplify the implementation of modules that
are executed by the cyclic executive, which itself is implemented in C.

See appendix A for the language reference.

\section{Runtime System}

The runtime system, including memory management, low-level message
passing, the cyclic executive scheduler, and an UART-driver for debugging output
consists of roughly 400 lines of C. {\it hO} is translated to C and linked with the
runtime system into an executable for final delivery to the processing unit
of the instrument. Note that all memory accesses are indirect -- exported
and internal data of modules is held in a module-specific static data block,
given to the module when it executes. That way module-data is position
independent and may be moved in case memory should fail due to radiation
damage. Moreover, all procedures are inlined, and no global variables are
used. This ensures all code is agnostic of the physical location of procedures
and variables on memory and can be moved at will.

Dynamic checks may result in error conditions that write debugging output to
the serial interface at development time or result in error event telemetry
and reboot (in case of fatal errors), when delivered in the flight model.
To make sure that error event handling takes place, regardless of the
state of the system, this mechanism has to circumvent the normal
event and telemetry handling.

Message memory is managed in a simple list of free 4 kB blocks.
There are basic operations for allocating a block ({\tt NEW}) and releasing
them back to the runtime system ({\tt DISPOSE}). All blocks have the same
size, and are passed from module to module, without copying. So, for example,
the SpaceWire\lref{5}\lref{6} module retrieves a packet from the hardware interface,
allocates and fills a block with the data and sends it to another module
responsible for validating the packet. This module in turn will send the
block further for more specific processing. Once the block/message/packet
has been fully processed, it can be released and given back to the runtime
system.

Direct access to the message contents is primitively only available on the
byte-level.
However, most telecommand/telemetry messages that SWI handles are structured
records as defined by ESTEC\lref{7}.
A machine-generated library of accessor functions is used to
take apart telecommands and construct telemetry packets, generated from
the TC/TM ({\it MIB}) database that has to be provided as a component of
the flight software.

\section{The Compiler}

The hO compiler is implemented in Scheme\lref{24} and parses our Oberon
dialect, performs type-checking and generates ANSI C99. The compiler
is implemented in a straightforward manner, using the {\it Silex}\lref{3}
lexer generator and a custom PEG\lref{9} parser. The type-checker implements
a straightforward static typing scheme and performs some additional checks that
we considered worthwhile. For documentation purposes the compiler
also provides generating a {\it dot} diagram depicting message-flow between
modules, using the {\it graphviz}\lref{13} tool, and generating support files for
mapping runtime error information to source code locations.

To allow circular module references, the compiler can be run in a restricted
mode that just generates interface information for each module, instead of
fully validating inter-module data access.

\section{Properties of the generated C code}

The use of hO eliminates many sources of undefined behaviour in C,
among these are:

\begin{enumerate}

\item NULL-pointer references

\item Pointer-arithmetic

\item Numeric overflow of signed integers, provided the necessary compiler
flags are given when compiling the generated C code

\item Right-shifts of signed integers always produce an unsigned result

\item Division/modulo by zero is detected and caught at runtime

\item Uninitialized variables

\end{enumerate}

The code generated by {\tt hoc} uses GNU extensions, namely
local (inline) functions and statement expressions. It is recommended
to compile the C code with the
{\tt -fno-strict-aliasing -fwrapv -std=gnu99}
compiler options to ensure correct semantics at runtime.

\section*{Appendix A -- hO Language Reference}
\addcontentsline{toc}{section}{Appendix A \\ hO Language Reference}

If not indicated differently, the lexical syntax of hO is the same as in
Oberon. Comments are written as

\begin{verbatim}
    (* ... *)
\end{verbatim}

and can not be nested.

hO is case-sensitive, builtin keywords may alternatively be given in
lowercase.

\subsection*{Toplevel forms}

\subsubsection*{Modules}

The application is structured into modules, where each module contains
some toplevel code, executed periodically, and a private and optionally
public set of global variables. Modules may be {\it instantiated} multiple times,
for example, to provide implementations for two redundant hardware
interfaces or for multiple variants of the same functionality. In SWI we
have a second redundant SpaceWire interface and both the spectrometer
and autocorrelator exist in 2 versions.

A module must return to its caller, or the system will grind to a halt, but
since there are no syntactic constructs to perform unbounded loops or
mutually recursive procedures, such a situation is not possible.

A module is defined using the syntax

\begin{verbatim}
    MODULE name;
    VAR exported*: u32, listener*: port;
           secret, unknown: u32;
    BEGIN
        ...
    END name.
\end{verbatim}

Here, {\tt exported} and {\tt listener} are externally visible, {\tt secret} and {\tt unknown}
are internal. The variables declared are visible in the scope of the module
and retain their
contents over every cyclic execution. To access the exported
variables of another module the {\tt name.listener} syntax can be used,
provided the module {\tt name} was imported into the current scope using
them {\tt IMPORT} form.

Module variables are always initialized to {\tt 0} (zero) or, in the
case of ``port'' objects, to empty ports.

Modules that exist in multiple intstances (currently at most 2) are defined
like this:

\begin{verbatim}
    MODULE spw*;
    ...
\end{verbatim}

The {\tt spw} module above will exist in two variants ({\tt spw0} and {\tt spw1})
in the final image. Note that though they do share source code,
at runtime they share neither code nor data.

To access the public variables of another module, use the {\tt IMPORT}
statement:

\begin{verbatim}
    IMPORT some_module;
\end{verbatim}

If a module provides several instances, then access to the variables of a
particular instance, or of the instance corresponding to the instance
of the current module can be specified as follows:

\begin{verbatim}
    IMPORT mod0 := mod[0];   (* only access instance 0 *)
    IMPORT mod := mod[*];   (* import corresponding instance *)
\end{verbatim}

{\tt mod}s variables can be accessed now using {\it dot} notation, as in

\begin{verbatim}
    SEND(msg, mod.port);
\end{verbatim}

\subsubsection*{Message passing}

Modules communicate by sending messages, A message is a raw {\tt u8}
array of at most 4096 bytes. Any variable of type {\tt port} can
hold at most one message or is empty. Sending a message from one port
to another port moves the data array: sending a message from an empty
port or to a blocked port does nothing, otherwise the target port
holds the message and the source port is empty.

Uninitialized ports are empty

\subsubsection*{Type-, constant and procedure definitions}

Types and constants are defined as in Oberon:

\begin{verbatim}
    TYPE point = RECORD x, y: s32 END;
    CONST zero = 0;
\end{verbatim}

Builtin types exist for basic numeric types: {\tt u8}, {\tt u16}, {\tt u32}, {\tt s8},
{\tt s16}, {\tt s32} and {\tt boolean}.

Procedure and function definitions are also similar to Oberon,
with the exception of unspecified {\tt ARRAY} arguments -- in hO all arrays
must have a fixed, statically known length. Procedures are always inlined,
so should be kept at a reasonable size.

Procedures may be declared at the start of a module to define internal
procedures, not visible in other modules. These internal procedures
have full access to the public and private module variables.

Procedures may modify their argument variables for arguments declared
with {\tt VAR} (as in Oberon.)

\subsubsection*{Contracts}

hO supports a basic Design-by-Contract facility to run dynamic checks
at runtime. A contract is a boolean function, optionally with arguments.
Contracts must be defined at toplevel and can not be local to a module.

\begin{verbatim}
    CONTRACT ensure_positive(x: s32)
    BEGIN
        return x > 0
    END;
\end{verbatim}

Contracts may not use {\tt VAR} arguments.

To use a contract, use the {\tt REQUIRE}, {\tt PROVIDE} and {\tt INVARIANT}
statements at the start of a statement sequence, that is, at the
start of a module-, procedure, conditional or loop body:

\begin{verbatim}
    LOCAL counter := 0;
    REPEAT 10 TIMES
        REQUIRE ensure_positive(counter), other_contract;
        ...
    END;
\end{verbatim}

Contracts used by {\tt REQUIRE} are checked at the start of the
respective statement sequence,
contracts used by {\tt PROVIDE} at the end and contracts used by
{\tt INVARIANT} are checked both at the start and at the end of
a sequence, respectively.

\subsubsection*{Include Files}

The contents of external files can be included using the {\tt INCLUDE} form:

\begin{verbatim}
    INCLUDE "some_file.ho";
\end{verbatim}

An include-file is included at most once. Including the same file subsequently
has no effect.

\subsection*{Statements}

The following statements are allowed:

\subsubsection*{Conditionals}

\begin{verbatim}
    IF expression THEN ... [ELSIF ...] [ELSE ...] END
    CASE expression OF const1: ... | const2: ... [ELSE ...] END
\end{verbatim}

{\tt IF} is the usual conditional statement. {\tt CASE} provides an n-way
comparison with optional default ({\tt ELSE}) clause. Multiple constants
and constant ranges can be used in {\tt CASE} clauses, as in Oberon.
If no case applies, then execution continues normally after the
{\tt CASE} construct. At the end of a clause, execution continues
after the {\tt CASE} statement.

\subsubsection*{Loops}

There is only a single loop statement available, a counted loop with
an optional condition:

\begin{verbatim}
    [WHILE expression] REPEAT constant TIMES ... END
\end{verbatim}

The loop counter must be constant to ensure termination.
The expression is evaluated before each loop iteration.
A dedicated
loop variable is not provided, but can be defined and updated manually.
Note that regardless of the value of the guard expression, the loop
will terminate after the indicated number of iterations.

\subsubsection*{State machines}

Builtin support for state machines is provided. You can declare states
using the {\tt STATE} form:

\begin{verbatim}
    STATE init, process, idle;
\end{verbatim}

{\tt STATE} defines each declared state as a constant of type {\tt s32},
starting with {\tt 0} (so an uninitialized state variable implicitly
holds the first state.)

The {\tt SELECT} statement dispatches between states, and {\tt NEXT}
switches to another state:

\begin{verbatim}
    SELECT current OF state1: ... | state2: ... END
\end{verbatim}

{\tt current} must be an assignable value or type {\tt u32}.
{\tt NEXT} assigns a new state to the state variable given:

\begin{verbatim}
    NEXT process;
\end{verbatim}

There is no ``fall-through'' in state clauses: at the end of the
statement-sequence in a state, execution commences after the
{\tt SELEC} statement.

Note that {\tt NEXT} does not perform a transfer of control to a different
state clause, but to the end of the select statement instead.
On the next execution of the select statement,
the state variable given in the {\tt SELECT} statement
will have the new value assigned to it in the previous iteration.

{\tt NEXT} is not allowed to exit a looping construct.

If no clause of the {\tt SELECT} form applies to the current state,
execution continues after the {\tt SELECT} form as normal.

\subsubsection*{Other statements}

hO supports the usual procedure calls, assignments and {\tt RETURN}
statements as in Oberon. Local variables may be declared at any
point using

\begin{verbatim}
    LOCAL var := 0;
\end{verbatim}

The scope of the variable extends to the current block, that is,
to the next outer {\tt END} token.
Local variables must be initialized and may optionally be declared to
have a particular type, if the type of the initialization expression is
insufficient:

\begin{verbatim}
    LOCAL unsigned := 0: u32;
\end{verbatim}

So called {\it external} variables allow accessing direct memory locations:

\begin{verbatim}
    EXTERNAL uart_reg := 80000100h: VOLATILE POINTER TO u32
\end{verbatim}

Should a locally defined variable shadow an existing variable
in an outer scope, then the compiler will issue a warning.

\subsection*{Expressions}

The usual arithmetic and comparison operators of Oberon are provided
in hO as well. Since hO does not support floating-point arithmetic, {\tt /}
and {\tt DIV} are equivalent. Consult the EBNF grammar later in this section
for information about operator precedence.

\subsection*{Builtin types, constants, operators and procedures}

\subsubsection*{Types}

\begin{list}{}{
  \setlength{\leftmargin}{6ex}
  \setlength{\itemindent}{-3ex}}

\item {\tt s8}, {\tt s16}, {\tt s32}, {\tt u8}, {\tt u16}, {\tt u32} \\
        Numerical base types.

\item {\tt boolean} \\
        The boolean type.

\item {\tt port} \\
        The type of a structure that can hold messages.

\item {\lbrack {\tt VOLATILE}\rbrack ~ {\tt POINTER} {\tt TO} {\it type}} \\
        A pointer type, optionally declared volatile, to disallow assumption made by
        the compiler about the value of memory locations, when the memory
        location might be updated asynchronously, for example.
\end{list}

\subsubsection*{Constants}

\begin{list}{}{
  \setlength{\leftmargin}{6ex}
  \setlength{\itemindent}{-3ex}}

\item {\tt MODULE\_ID}: {\tt u32} \\
        A constant holding the numeric ID of the current module.

\item {\tt INSTANCE}: {\tt u32} \\
        A constant holding the instance-ID of the current module, if this module
        has several instances (currently 0 or 1).

\item {\tt TRUE}, {\tt FALSE}: {\tt boolean} \\
        The canonical true and false values.
\end{list}

\subsubsection*{Operators}

\begin{list}{}{
  \setlength{\leftmargin}{6ex}
  \setlength{\itemindent}{-3ex}}

\item {\tt AND} {\tt OR} {\tt NOT} \\
        Boolean operators. Note that {\tt AND} and {\tt OR} are short-circuiting as in C.

\item {\tt /\textbackslash} {\tt \textbackslash/} {\tt ><} {\tt \textasciitilde} \\
        Binary logic operators (and, or, xor, not).

\item {\tt >>} {\tt <<} \\
        Perform unsigned left or right shift of {\it x} by {\it y} bits.

\item {\tt +} {\tt -} {\tt *} {\tt /} {\tt DIV} {\tt MOD} \\
        Numerical operators. Arithmetic between arguments of differing numeric
        type are allowed. {\tt DIV} is currently an alias
        for {\tt /}.

\item {\tt =} {\tt \#} {\tt >} {\tt <} {\tt >=} {\tt <=} \\
        Comparison operators for arguments of numeric type.

\item {\tt SIZE(}{\it type}{\tt )} \\
        Returns the size in bytes of the memory occupied by {\it type}.
\end{list}

Expressions of boolean or comparison operators that produce a
constant result will produce a warning as this may indicate a
programming error.

Operator precedence is as follows, listed from highest precedence
to lowest, sequent binary operators are evaluated from left to right:

\begin{enumerate}
\item Selection operators ({\tt . [] \textasciicircum})
\item Subexpression ({\tt ()})
\item Unary operators (unary {\tt + - NOT \textasciitilde} and {\tt SIZE})
\item Multiplicative operators ({\tt * / DIV MOD AND /\textbackslash})
\item Shift operators ({\tt >> <<})
\item Additive operators (binary {\tt + -} and {\tt OR \textbackslash/ ><})
\item Comparison operators ({\tt = \# <= => < >})
\end{enumerate}
\subsubsection*{Procedures}

\begin{list}{}{
  \setlength{\leftmargin}{6ex}
  \setlength{\itemindent}{-3ex}}

\item {\tt CLONE(}{\it oldport}{\tt ,}{\it newport}{\tt )} \\
        Clones the message stored by {\it oldport} and puts a copy into {\it newport},
        both of which must be port objects. {\it oldport} may be empty.

\item {\tt DEC(}{\it var}{\tt )} \\
\hspace*{-3.7ex} {\tt INC(}{\it var}{\tt )} \\
        Increase or decrease the numeric value in location {\it var}.

\item {\tt DISPOSE(}{\it port}{\tt )} \\
        Free the memory used by the message in {\it port}.
        If the {\it port} is empty, do nothing.

\item {\tt EXTEND(}{\it port}{\tt ,}{\it n}{\tt )} \\
        Increase or decreases the used size of the message in {\it port} by {\it n}. {\tt n} may be negative.

\item {\tt LOG(}{\it string}{\tt )} \\
\hspace*{-3.7ex} {\tt LOG(}{\it string}{\tt ,}{\it x}{\tt )} \\
        Writes debugging output to the serial interface, optionally with a numeric
        argument.

\item {\tt NEW(}{\it port}{\tt ,}{\it size}{\tt )} \\
        Allocates a buffer of the given {\it size} for holding a message and stores it in {\it port}. The message is uninitialized and contains
random data.

\item {\tt SEND(}{\it fromport}{\tt ,}{\it toport}{\tt )} \\
        Moves a message from port to port. If the receiver port is non-empty, this
        operation is a no-op. If the receiver port is empty, the sender port
        will be emptied.
\end{list}

\subsubsection*{Functions}

\begin{list}{}{
  \setlength{\leftmargin}{6ex}
  \setlength{\itemindent}{-3ex}}

\item {\tt ADR(}{\it var}{\tt ): POINTER TO}{\it type} \\
        Returns a pointer to the location given by {\it var}.

\item {\tt COUNT(}{\it port}{\tt ): s32} \\
        Returns the size of the message stored in {\it port}, in bytes.

\item {\tt DATA(}{\it port}{\tt ): ARRAY} {\it blocksize} {\tt OF u8} \\
        Returns the data array holding the data of any message in {\it port}.

\item {\tt MAX(}{\it x}{\tt ,}{\it y}{\tt ): s32} \\
\hspace*{-3.7ex} {\tt MIN(}{\it x}{\tt ,}{\it y}{\tt ): s32} \\
        The minimum and maximum functions.

\item {\tt PENDING(}{\it port}{\tt ): boolean} \\
        Returns true if {\it port} holds a message.

\item {\tt SEND(}{\it fromport}{\tt ,}{\it toport}{\tt ): boolean} \\
        As the {\tt SEND} procedure, but returns a boolean, indicating
       whether the message was actually sent.
\end{list}

\subsection*{EBNF Grammar}

\noindent
$compilation$-$unit ::=$
  \{ $toplevel$ ``{\tt ;}'' \} .\\
$toplevel ::=$
  $typedef$
  $\vert$ $module$
  $\vert$ $function$
  $\vert$ $contract$
  $\vert$ $include$
  $\vert$ $constdef$ .\\
$include ::=$
  ``{\tt INCLUDE}'' $string$ .\\
$constdef ::=$
  ``{\tt CONST}'' \{ $id$ ``{\tt =}'' $constant$ ``{\tt ;}'' \} .\\
$typedef ::=$
  ``{\tt TYPE}'' \{ $id$ ``{\tt =}'' TYPE ``{\tt ;}'' \} .\\
$function ::=$
  ``{\tt PROCEDURE}'' $id$ ``{\tt (}'' $param$ \{ ``{\tt ,}'' $param$ \} ``{\tt )}'' $\lbrack$ ``{\tt :}'' $type$ $\rbrack$\\
\hspace*{8ex}
  ``{\tt BEGIN}'' $statements$ ``{\tt END}'' .\\
$contract ::=$
  ``{\tt CONTRACT}'' $id$ $\lbrack$ ``{\tt (}'' $param$ \{ ``{\tt ,}'' $param$ \} ``{\tt )}'' $\rbrack$\\
\hspace*{8ex}
  ``{\tt BEGIN}'' $statements$ ``{\tt END}'' .\\
$param ::=$
  $\lbrack$ ``{\tt VAR}'' $\rbrack$ $id$ \{ ``{\tt ,}'' $id$ \} ``{\tt :}'' $type$ .\\
$statement ::=$
  $if$
  $\vert$ $loop$
  $\vert$ $return$
  $\vert$ $pcall$
  $\vert$ $variable$
  $\vert$ $external$
  $\vert$\\
\hspace*{8ex}
  $statedef$
  $\vert$ $import$
  $\vert$ $select$
  $\vert$ $log$
  $\vert$ $next$
  $\vert$ $case$
  $\vert$ $assign$
  $\vert$ ``{\tt ;}'' .\\
$statements ::=$
  \{ $check$ ``{\tt ;}'' \} $statement$ \{ ``{\tt ;}'' $statements$ \} .\\
$log ::=$
  ``{\tt LOG}'' ``{\tt (}'' $string$ $\lbrack$ ``{\tt ,}'' $expr$ $\rbrack$ ``{\tt )}'' .\\
$check ::=$
  ``{\tt REQUIRE}'' $assert$ \{ ``{\tt ,}'' $assert$ \} $\vert$\\
\hspace*{8ex}
  ``{\tt PROVIDE}'' $assert$ \{ ``{\tt ,}'' $assert$ \} $\vert$\\
\hspace*{8ex}
  ``{\tt INVARIANT}'' $assert$ \{ ``{\tt ,}'' $assert$ \} .\\
$assert ::=$
  $pcall$ $\vert$ $id$ .\\
$variable ::=$
  ``{\tt LOCAL}'' $id$ ``{\tt :=}'' $expr$ $\lbrack$ ``{\tt :}'' $type$ $\rbrack$\\
\hspace*{8ex}
  \{ ``{\tt ,}'' $id$ ``{\tt :=}'' $expr$ $\lbrack$ ``{\tt :}'' $type$ $\rbrack$ \} .\\
$external ::=$
  ``{\tt EXTERNAL}'' $id$ ``{\tt :=}'' $constant$ ``{\tt :}'' $type$\\
\hspace*{8ex}
  \{ ``{\tt ,}'' $id$ ``{\tt :=}'' $constant$ ``{\tt :}'' $type$ \} .\\
$statedef ::=$
  ``{\tt STATE}'' $id$ \{ ``{\tt ,}'' $id$ \} .\\
$import ::=$
  ``{\tt IMPORT}'' $importspec$ \{ ``{\tt ,}'' $importspec$ \} .\\
$importspec ::=$
  $id$ $\lbrack$ ``{\tt :=}'' $id$ $\lbrack$ ``{\tt [}'' ( ``{\tt *}'' $\vert$ $constant$ ) ``{\tt ]}'' $\rbrack$  $\rbrack$ .\\
$select ::=$
  ``{\tt SELECT}'' $lvalue$ ``{\tt OF}'' $state$ \{ ``{\tt $\vert$}'' $state$ \} ``{\tt END}'' .\\
$state ::=$
  $range$ ``{\tt :}'' $statements$ .\\
$range ::=$
  $range$-$element$ \{ ``{\tt ,}'' $range$-$element$ \} .\\
$range$-$element ::=$
  $constant$ $\lbrack$ ``{\tt ..}'' $constant$ $\rbrack$ .\\
$next ::=$
  ``{\tt NEXT}'' $constant$ .\\
$loop ::=$
  $\lbrack$ ``{\tt WHILE}'' $expr$ $\rbrack$ ``{\tt REPEAT}'' $constant$ ``{\tt TIMES}'' $statements$ ``{\tt END}'' .\\
$case ::=$
  ``{\tt CASE}'' $expr$ ``{\tt OF}'' $state$ \{ ``{\tt $\vert$}'' $state$ \} $\lbrack$ ``{\tt ELSE}'' $statements$ $\rbrack$ ``{\tt END}'' .\\
$assign ::=$
  $designator$ ``{\tt :=}'' $expr$ .\\
$return ::=$
  ``{\tt RETURN}'' $\lbrack$ $expr$ $\rbrack$ .\\
$if ::=$
  ``{\tt IF}'' $expr$ ``{\tt THEN}'' $statements$ \{ ``{\tt ELSIF}'' $statements$ \}\\
\hspace*{8ex}
  $\lbrack$ ``{\tt ELSE}'' $statements$ $\rbrack$ ``{\tt END}'' .\\
$pcall ::=$
  $id$ ``{\tt (}'' $expr$ \{ ``{\tt ,}'' $expr$ \} ``{\tt )}'' .\\
$expr ::=$
  $sum$ $\lbrack$ ( ``{\tt =}'' $\vert$ ``{\tt \#}'' $\vert$ ``{\tt <=}'' $\vert$ ``{\tt >=}'' $\vert$ ``{\tt <}'' $\vert$ ``{\tt >}'' ) $sum$ $\rbrack$ .\\
$sum ::=$
  $term$ \{ ( ``{\tt +}'' $\vert$ ``{\tt -}'' $\vert$ ``{\tt OR}''
  $\vert$ ``{\tt \textbackslash/}'' $\vert$ ``{\tt ><}'' ) $term$ \} .\\
$term ::=$
  $product$ $\lbrack$ ( ``{\tt <<}'' $\vert$ ``{\tt >>}'' ) $product$ $\rbrack$ .\\
$product ::=$
  $unary$ \{ ( ``{\tt *}'' $\vert$ ``{\tt /}'' $\vert$ ``{\tt DIV}'' $\vert$ ``{\tt MOD}''
  $\vert$ ``{\tt AND}'' $\vert$ ``{\tt /\textbackslash}'' ) $unary$ \} .\\
$unary ::=$
  $\lbrack$ ``{\tt +}'' $\vert$ ``{\tt -}'' $\vert$ ``{\tt \textasciitilde}'' $\vert$ ``{\tt NOT}'' $\rbrack$ $factor$
  $\vert$ ``{\tt SIZE}'' ``{\tt (}'' $type$ ``{\tt )}'' .\\
$factor ::=$
  $designator$ $\vert$ $number$ $\vert$ ``{\tt (}'' $expr$ ``{\tt )}'' $\vert$ $boolean$ .\\
$boolean ::=$
  ``{\tt TRUE}'' $\vert$ ``{\tt FALSE}'' .\\
$designator ::=$
  ( $id$ $\vert$ $pcall$ ) \{ ``{\tt .}'' $id$ $\vert$ ``{\tt [}'' $expr$ ``{\tt ]}'' $\vert$ ``{\tt \^{}}'' \} .\\
$module ::=$
  ``{\tt MODULE}'' $id$ $\lbrack$ ``{\tt *}'' $\rbrack$ ``{\tt ;}'' $\lbrack$ modulevars $\rbrack$ $\lbrack$ modulefunctions $\rbrack$\\
\hspace*{8ex}
  ``{\tt BEGIN}'' $statements$ ``{\tt END}'' $id$ ``{\tt .}'' .\\
$modulevars ::=$
  ``{\tt VAR}'' $vardef$ \{ ``{\tt ;}'' $vardef$ \} .\\
$modulefunctions ::=$
  $function$ \{ ``{\tt ;}'' $function$ \} .\\
$vardef ::=$
  $id$ $\lbrack$ ``{\tt *}'' $\rbrack$ \{ ``{\tt ,}'' $id$ $\lbrack$ ``{\tt *}'' $\rbrack$ \} ``{\tt :}'' $type$ .\\
$type ::=$
  $simpletype$
  $\vert$ $structtype$
  $\vert$ $ptrtype$
  $\vert$ $arraytype$ .\\
$ptrtype ::=$
  $\lbrack$ ``{\tt VOLATILE}'' $\rbrack$ ``{\tt POINTER}'' ``{\tt TO}'' $type$ .\\
$structtype ::=$
  ``{\tt RECORD}'' $field$ \{ ``{\tt ;}'' $field$ \} ``{\tt END}'' .\\
$field ::=$
  $id$ \{ ``{\tt ,}'' $id$ \} ``{\tt :}'' $type$ .\\
$arraytype ::=$
  ``{\tt ARRAY}'' $constant$ ``{\tt OF}'' $type$ .\\
$simpletype ::=$
  $id$ .\\
$constant ::=$
  $expr_{const}$ .\\
$id ::=$
  ($letter$$\vert$``{\tt \_}'')\{$letter$$\vert$``{\tt \_}''$\vert$$digit$\}.\\
$string ::=$
  {\tt "}\{$character$\}{\tt "} .\\
$number ::=$
  $digit$\{$digit$\}$\lbrack$\{$hexdigit$\}``{\tt h}''$\rbrack$ $\vert$ {\tt '}\{$character\}${\tt '} .\\

\section*{Appendix B -- Compiler Usage}
\addcontentsline{toc}{section}{Appendix B \\ Compiler Usage}

If correctly installed, the hO$\rightarrow$C translator can be invoked by entering
the {\tt hoc} command, passing the name of a hO source file and zero or
more of the following command line options:

\begin{list}{}{
  \setlength{\leftmargin}{6ex}
  \setlength{\itemindent}{-3ex}}

\item
    {\tt -h} \\
        Show a short message listing the available options

\item
    {\tt -I} {\it directory} \\
        Specifies an additional include {\it directory} to be added to the search
        path for files accessed via the {\tt INCLUDE} form.

\item
    {\tt -d} \\
        Generate dependency rules for {\tt make(1)}, writing them to the standard
        output channel.

\item
    {\tt -f} \\
        Generate a message flow diagram suitable for the {\tt dot(1)} tool, writing
        it to the standard output channel.

\item
    {\tt -o} {\it filename} \\
        Specify an alternative output {\it filename}. The default is to use the name
        of the source file, but with the `{\tt .c}' extension.

\item
    {\tt -g} \\
        Only generate an interface file, named after the source filename, but with
        the `{\tt .hi}' extension.

\item
    {\tt -k} \\
        Keep intermediate files from the parsing, processing and type-checking
        stages.
\end{list}

\section*{Appendix C -- Dynamic runtime checks}
\addcontentsline{toc}{section}{Appendix C \\ Dynamic runtime checks}

The following conditions are dynamically checked at runtime, unless
being explicitly disabled by compiling the C files generated
by {\tt hoc} with the {\tt -DNDEBUG} option:

\begin{list}{}{
  \setlength{\leftmargin}{6ex}
  \setlength{\itemindent}{-3ex}}

\item Empty port access \\
    Accessing the {\tt DATA} of an empty port.

\item Division/modulo by zero \\
    Using {\tt /}, {\tt DIV} or {\tt MOD} with a second argument
    that is zeri.

\item Invalid shift \\
    Using {\tt <<} or {\tt >>} with a second argument that is
    negative or equal or higher than the word size of the
    used processor architecture.

\item Array bounds \\
    Accessing an array with an index that is negative or exceeds
    the size of the array.

\end{list}

\section*{Appendix D -- Proposed coding style}
\addcontentsline{toc}{section}{Appendix D \\ Proposed coding style}

The following is a description of the coding style that we use in
the JUICE~SWI project. We recommend to follow this style when
using hO.

\begin{enumerate}

\item A line of code should contain at most a single statement. If
the statement is a complete control structure like {\tt IF/THEN} and
the body contains a single statements that fits into the line, then
the conditional may occupy the complete line.
\item {\tt THEN} and {\tt OF} should terminate
 a line, possibly with a trailing semicolon {\tt ;}).
\item {\tt ELSE} and {\tt END} should always be on a separate line.
\item Define local or external variables as near to their first use
 as possible.
\item Declare local variables to a specific type to avoid expression ambiguity.
\item Avoid complex expressions that exceed a single line.
\item Indent {\tt CASE} and {\tt SELECT} cases consistently.
\item Separate {\tt VAR} blocks and module bodies by an empty line.
\item Surround binary expression- and assignment operators with a single space on both sides.
\item Separate toplevel entities (function- and module definitions, also
comment blocks)
 by at least one empty line. Otherwise avoid empty lines in source code.
\item Avoid unnecessary comments unless they convey something
 important or critical to the understanding of the code.
\item Don't use source lines that exceed 100 columns.
\item Write keywords in lowercase.
\item Avoid TAB characters as there is no standard width of a TAB
 and source code may look differently depending on client settings.
\item Whatever you do in following or deviating from this coding style,
 be consistent.
\end{enumerate}

\subsection*{References}

\def\Lit#1#2{\item {\tt [#1]}
#2}
\begin{list}{}{
  \setlength{\labelwidth}{0mm}
  \setlength{\itemsep}{0ex plus0.2ex}
  \setlength{\leftmargin}{6ex}
  \setlength{\itemindent}{-6ex}
  \setlength{\labelsep}{0mm}}

\Lit{1}{
  Baker, T. P. and Shaw, A.:
  ``The cyclic executive model and Ada'',
  The Journal of Real-Time Systems~1, 1989, pp.~7-25
}

\Lit{2}{
  E. Brewer et al:
  ``Thirty Years is Long Enough: Getting Beyond C.'',
  Proceedings of the 10th conference on
  Hot Topics in Operating Systems -- Volume 10,
  USENIX Association, 2005
}

\Lit{3}{
  Danny Dub\'{e}:
  ``A Scheme Implementation of Lex'',
  2001
}

\Lit{4}{
  Elliot, Seidel, Launchbury et al:
  ``Guilt Free Ivory'',
  Haskell Symposium '15
}

\Lit{5}{
  ECSS Secretariat:
  ``Space engineering -- SpaceWire protocol identification''
  (ECSS-E-ST-50-51C),
  ESA-ESTEC, Noordw{\ij}k, 5.2.2010
}

\Lit{6}{
  ECSS Secretariat:
  ``SpaceWire -- Links, nodes, routers and networks''
  (ECSS-E-ST-50-12C),
  ESA-ESTEC, Noordw{\ij}k, 31.7.2008
}

\Lit{7}{
  ECSS Secretariat:
  ``Space engineering -- Telemetry and telecommand packet utilization''
  ~ (ECSS-E-ST-70-41C),
  ~ ESA-ESTEC, ~ Noordw{\ij}k, 15.4.2016
}

\Lit{8}{
  ``JUICE -- Exploring the emergence of habitable worlds around gas giants''
  (ESA/SRE(2011)18), December 2011,
  European Space Agency
}

\Lit{9}{
  Bryan Ford:
  ``Parsing expression grammars: a recognition-based syntactic foundation'',
  Proceedings of the 31st ACM SIGPLAN-SIGACT symposium on Principles of programming languages,
  p.111-122,
  Venice, 2004
}

\Lit{10}{
  Aeroflex Gaisler AB:
  ``LEON3-FT SPARC V8 Processor -- Data Sheet and User's Manual'',
  January 2013, Version 1.9
}

\Lit{11}{
  Cobham Gaisler AB:
  ``Bare-C Cross-Compiler -- User's Manual'',
  September 2016, Version 1.0.46
}

\Lit{12}{
  Cobham Gaisler AB:
  ``LEON3FT Stale Cache Entry After Store with Data Tag Parity Error''
  (Technical note GRLIB-TN-0009),
  September 2016, Issue 1.0
}

\Lit{13}{
  Emden R. Gansner, Stephen C. North:
  ``An open graph visualization system and its applications to software engineering'',
  1999
}

\Lit{14}{
  P. Hartogh et al:
  ``The Submillimetre Wave Instrument on JUICE'',
  European Planetary Science Congress 2013,
  Vol. 8, EPSC2013-710, 2013
}

\Lit{15}{
  Gerard J. Holzmann:
  ``The Power of Ten -- Rules for Developing Safety Critical Code'',
  2006,
  NASA/JPL Laboratory for Reliable Software,
  Pasadena, CA 91109
}

\Lit{16}{
  ``Ada Reference Manual'',
  ISO/IEC 8652:1995(E)
}

\Lit{17}{
  Brian W. Kernighan, Dennis M. Ritchie:
  ``The C Programming Language (1st ed.)'',
  February 1978,
  Englewood Cliffs, NJ: Prentice Hall, ISBN 0-13-110163-3
}

\Lit{18}{
  C. Douglass Locke:
  ``Software Architecture for Hard Real-Time Applications: Cyclic Executive vs. Fixed Priority Executives'',
  The Journal of Real-Time Systems~4, 1992, pp.~37-53
}

\Lit{19}{
  Meyer, Bertrand:
  ``Design by Contract'', Technical Report TR-EI-12/CO,
  Interactive Software Engineering Inc., 1986
}

\Lit{20}{
  ``MISRA-C:2004 -- Guidelines for the use of the C language in critical systems'',
  MIRA Ltd.,
  October 2004
}

\Lit{21}{
  Moore, C.H.:
  ``Programming a Problem-Oriented Language'',
  Amsterdam, NY: Mohasco Industries Inc. (internal pub.) 1970
}

\Lit{22}{
  Oskar Schirmer:
  ``GuStL -- An Experimental Guarded States Language'',
  G\"{o}ttingen, 2016
}

\Lit{23}{
  Paul Soulier, Depeng Li:
  ``Blueprint for an Embedded Systems Programming Language''
}

\Lit{24}{
  Gerald Jay Sussman and Guy Lewis Steele, Jr.:
  ``Scheme: An Interpreter for Extended Lambda Calculus'',
  MIT AI Lab.,
  AI Lab Memo AIM-349,
  December 1975
}

\Lit{25}{
  ``CHICKEN -- A Practical and Portable Scheme System'', \\
  http://www.call-cc.org
}

\Lit{26}{
  Niklaus Wirth: ~
  ``The Programming Language Oberon'', ~
  Revision 1.10.2013~/~3.5.2016
}

\end{list}

\end{document}